\documentclass{article}

\usepackage{amsmath}  % load first
\usepackage[english]{babel}
\usepackage[utf8]{inputenc}
\usepackage{amssymb}
\usepackage{graphicx}
\usepackage{authblk}
\usepackage{caption}
\usepackage{multirow}
\usepackage{amscd}
\usepackage{geometry}
\usepackage{a4wide}
\usepackage[usenames]{color}
\usepackage{etaremune} 
\usepackage[hidelinks]{hyperref}
\usepackage{amsfonts}
\usepackage{ifthen}
\usepackage{color}
\usepackage{amscd}
\usepackage{latexsym}
\usepackage{tikz-cd}

\newcommand{\bd}{\begin{definition}}
	\newcommand{\ed}{\end{definition}}
\newcommand{\bt}{\begin{theorem}}
	\newcommand{\et}{\end{theorem}}
\newcommand{\bi}{\begin{itemize}}
	\newcommand{\ei}{\end{itemize}}
\newcommand{\ben}{\begin{enumerate}}
	\newcommand{\een}{\end{enumerate}}
\newcommand{\beq}{\begin{equation}}
\newcommand{\eeq}{\end{equation}}

\newtheorem{definition}{Def.}[section]
\newtheorem{theorem}{Theorem}[section]

\begin{document}

\title{Split-quaternions for perceptual white balance}

\author[1]{Michel Berthier\thanks{michel.berthier@univ-lr.fr}}
\author[2]{Nicoletta Prencipe\thanks{nicoletta.prencipe@oulu.fi}}
\author[3]{Edoardo Provenzi\thanks{edoardo.provenzi@math.u-bordeaux.fr}}
%\author[1]{Author B\thanks{B.B@university.edu}}
%\author[1]{Author C\thanks{C.C@university.edu}}
%\author[2]{Author D\thanks{D.D@university.edu}}
%\author[2]{Author E\thanks{E.E@university.edu}}
\affil[1]{Laboratoire MIA, Bâtiment Pascal, Pôle Sciences et Technologies, Université de La Rochelle, 23, Avenue A. Einstein, BP 33060, 17031 La Rochelle Cedex, France}
\affil[2]{Center for Ubiquitous Computing, Faculty of Information Technology and Electrical Engineering, University of Oulu, Oulu, Finland}
\affil[3]{Université de Bordeaux, CNRS, Bordeaux INP, IMB, UMR 5251\\ F-33400, 351 Cours de la Libération, Talence, France}

\renewcommand\Authands{ and }
\date{}

\maketitle

%\begin{history}
%\received{(Day Month Year)}
%\revised{(Day Month Year)}
%\end{history}

\begin{abstract}
We propose a perceptual chromatic adaptation transform for white balance that makes use of split-quaternions. The novelty of the present work, which is motivated by a recently developed quantum-like model of color perception, consists at stressing the link between the algebraic structures appearing in this model and a certain sub-algebra of the split-quaternions.
We show the potentiality of this approach for color image processing applications by proposing a 
chromatic adaptation transform, implemented via an appropriate use of the split-quaternion multiplication. 
Moreover, quantitative comparisons with the widely used state-of-the art von Kries chromatic adaptation transform are provided.
\end{abstract}

%\keywords{Schr\"{o}dinger's axioms of color space; Stiles' measure; Homogeneous spaces; Lie groups and algebras. MSC codes: 57S25, 57S99, 78-02}

\section{Introduction}%{Motivation and outline of the paper}
The main objective of this work is to describe a new white balance algorithm that is implemented by means of split-quaternions. The peculiarity of this algorithm is the fact that it is designed to fit coherently with a recently developed mathematical model of color perception \cite{BerthierProvenzi:2022PRS,Berthier:22}. This model provides an alternative to the CIE (Commission Internationale de l'\'Eclairage) description of colors by means of three coordinates in a colorimetric space, e.g. RGB, HSV, CIELab and so on. It also emphasizes the fact that a perceived color should be described as the result of a (perceptual) measurement procedure. The measurement equation, that is the cornerstone of the proposed algorithm, uses tools from quantum information and expresses the result of a so-called Lüders operation.

A complete mathematical description of this new paradigm about color perception is out of the scope of this work. For the sake of self-consistency, the essential concepts of this model will be recalled in section \ref{Perception}, the reader interested in more details can consult the following papers \cite{BerthierProvenzi:2022PRS,Berthier:22,Berthier:2020,Berthier:21JMIV,Berthier:2021JofImaging,Berthier:21JMP}. We deem worthwhile to mention that that this model permits to: intrinsically reconcile trichromacy with Hering’s opponency \cite{Berthier:2020,Berthier:21JMIV};
formalize Newton’s chromatic disk \cite{Berthier:2020}; single out the Hilbert-Klein hyperbolic metric as a natural perceptual chromatic distance \cite{Berthier:21JMP}; solve the long-lasting problem of bounding the infinite perceptual color cone to a convex finite-volume solid of perceived colors \cite{Berthier:2020,BerthierProvenzi:2022PRS}; predict uncertainty relations for chromatic opposition \cite{Berthier:2021JofImaging} and give coherent mathematical definitions of perceived color perceptual attributes \cite{Berthier:22}.

As we will underline with more detail in section \ref{Perception}, the color measurement equation takes place in the algebra $\mathcal{H}(2,\mathbb{R})$ of $2\times 2$ symmetric matrices with real entries. To obtain a meaningful description of the measurement process, this algebra should not be considered as the associative, non-commutative algebra w.r.t. the matrix product, but rather as a non-associative, commutative algebra w.r.t. the product given by the symmetrized matrix product. Such a non-associative, commutative algebra is called a Jordan algebra, see e.g. \cite{Baez:12} for more details.

As we will see in the following, three distinct incarnations of a certain suitable Jordan algebra $\mathcal{A}$ give rise to three different perspectives on color measurements: 

\begin{enumerate}
	\item $\mathcal A$ viewed as $\mathcal{H}(2,\mathbb{R})$, gives a quantum information point of view, where color measurements are expressed via the so-called L\"uders operations;
	\item $\mathcal A$ viewed as the direct sum $\mathbb{R}\oplus \mathbb{R}^2$ provides a geometric and relativistic interpretation, where color measurements are expressed via normalized Lorentz boosts in the 3-dimensional Minkowski space;
	\item $\mathcal A$ viewed as a sub-algebra $\mathbb{S}_0$ of the split-quaternion algebra $\mathbb S$ brings a purely algebraic point of view, in which simple algebraic operations on split-quaternions allow us to encode color measurements.
\end{enumerate}

The three representations of $\mathcal{A}$ are linked by explicit Jordan algebra isomorphisms, which allow us to pass from one to another and to obtain a simple formula for the measurement process by means of split-quaternions. It is precisely this formula that can be efficiently implemented in the white balance algorithm. Although understanding all these mathematical arguments is not necessary in practice to perform experiments, and maybe readers more interested in the computational aspects would prefer to skip them and to focus on Sec. \ref{Split} and \ref{sec:experiments}, we deem important to underline that the proposed algorithm relies on a rigorous mathematical modeling of color perception.

The outline of the paper is the following: in  Sec. \ref{Perception} we start by introducing %We introduce in Sec. \ref{Perception}
the basic definitions and notations necessary to formulate the measurement equation that defines a color perceived by an observer from a visual scene. We also explain how color measurements can be interpreted geometrically by means of Lorentz boosts. Sec. \ref{Split} is devoted to the split-quaternion point of view. After recalling some preliminary notions, we show how to encode the measurement equation by a so-called `sandwich formula'. We also make explicit the geometric operations in $\mathbb R^4$ corresponding to this last formula to explain the difference with the usual sandwich formula involving the split-quaternion conjugacy. Experiments with the white balance algorithm are presented in Sec. \ref{sec:experiments}. The practical implementation is detailed through appropriate approximations of a suitable subset of $\mathbb{S}_0$ used in the measurement equation. Comparisons with the well-known von Kries algorithm are also presented.

\medskip

\section{Color perception and color measurements}
\label{Perception}
As said in the introduction, the proposed algorithm for white balance relies on a measurement equation that describes, from a quantum information point of view, the color perceived by an observer from a visual scene. The purpose of this section is to introduce the basic definitions and notations that permit to describe this equation. For additional information the interested reader may consult the references mentioned in the introduction.

\subsection{Color measurements and quantum information}

$\mathcal{H}(2,\mathbb{R})$, endowed with the Jordan product $A\circ B = (AB+BA)/2$, is a commutative, but not associative, Jordan algebra
%\begin{equation}\label{eq:symmatrix}
%   \small{A\circ B = \frac{1}{2}(AB+BA).}
%\end{equation}
with domain of positivity 
$\overline{\mathcal H^+}(2,\mathbb R)$, given by the set of real positive semi-definite $2\times 2$ matrices. As usual in quantum information, mesurements are described by the duality between states and effects. In the context of color perception, the state space is that of a rebit, the real analog of a qubit, with the complex vector space $\mathbb C^2$ replaced by the real vector space $\mathbb R^2$. In \cite{BerthierProvenzi:2022PRS,Berthier:22} it is explained why this quantum structure, that emerges from the sole axiomatic approach, is perfectly adapted to translate mathematically Hering's color opponency. A so-called chromatic state $\textbf{s}$ is represented by a density matrix $\rho_{\textbf{s}}$:
\begin{equation}\label{eq:rho}
	\rho_{\textbf{s}}= \frac{1}{2}\begin{pmatrix} 1+s_1 & s_2 \\s_2 &1- s_1  \end{pmatrix},
\end{equation}
with $1-s_1^2-s_2^2\geq0$, or equivalently by a chromatic vector $\textbf{v}_{\textbf{s}}=(s_1,s_2)$, with $||\textbf{v}_{\textbf{s}}||\leq 1$. 
%The chromatic state $\textbf{s}$ is said to be pure if $||\textbf{v}_{\textbf{s}}||=1$, mixed otherwise, see \cite{Berthier:2020}. 
An effect \textbf{e} is represented by a matrix $\eta_{\textbf{e}}$: 
\begin{equation}\label{eq:effectmat}
	\eta_{\textbf{e}}= \begin{pmatrix}
		e_0 + e_1 & e_2 \\ e_2 & e_0-e_1
	\end{pmatrix}= e_0\begin{pmatrix}
		1 + e_1/e_0 & e_2/e_0 \\ e_2/e_0 & 1-e_1/e_0
	\end{pmatrix},
\end{equation}
such that ${\bf 0}\le \eta_{\textbf{e}} \le Id_2$, or equivalently by a chromatic vector $\textbf{v}_{\textbf{e}}=(e_1/e_0, e_2/e_0)$ with $||\textbf{v}_{\textbf{e}}||\leq1$ and $0\leq e_0 \leq 1$. Given an effect \textbf{e}, associated to a human observer, and a chromatic state \textbf{s}, associated to the preparation of a visual scene, we have that the color perceived, or measured, by \textbf{e} from \textbf{s} is encoded in the outcome of the so-called L\"uders operation parametrized by $\textbf{e}$ and acting on $\textbf{s}$, explicitly \cite{Busch:97}:
\begin{equation}\label{eq:fund}
	\psi_{\textbf{e}}(\textbf{s}) = \eta_{\textbf{e}}^{1/2}\rho_{\textbf{s}}\eta_{\textbf{e}}^{1/2}.
\end{equation}
This last formula is the main topic of interest of this work. It is shown in \cite{Berthier:22} how it naturally permits to derive a coherent system of mathematical definitions of the CIE perceptual attributes, such as lightness, brightness, saturation and hue, using quantum information tools such as relative entropy.

%where $\eta_{\textbf{e}}^{1/2}$ is the so-called Kraus operator and $\langle \textbf{e} \rangle_{\textbf{s}}=\text{Tr}(\psi_{\textbf{e}}(\textbf{s}))$.

\subsection{Color measurements and Lorentz boosts}
The commutative Jordan product of the so-called spin factor $\mathbb{R}\oplus\mathbb{R}^2$ is given by $ (\alpha, \textbf{v})\circ (\beta, \textbf{w}) = (\alpha\beta + \langle \textbf{v},\textbf{w}\rangle, \alpha\textbf{w}+\beta\textbf{v})$
%\begin{equation}
%\footnotesize{ (\alpha, (v_1,v_2))\circ (\beta, (w_1,w_2)) = (\alpha\beta + v_1w_1 + v_2w_2, \alpha(w_1,w_2)+\beta(v_1,v_2)).}
%\end{equation}
%\begin{equation}
% (\alpha, \textbf{v})\circ (\beta, \textbf{w}) = (\alpha\beta + \langle \textbf{v},\textbf{w}\rangle, \alpha\textbf{w}+\beta\textbf{v}),
% \end{equation}
with $\alpha,\beta\in \mathbb{R}$, $\textbf{v},\textbf{w}\in \mathbb{R}^2$. The domain of positivity of $\mathbb{R}\oplus \mathbb{R}^2$ is $\overline{\mathcal L^+}=\{(\alpha,{\bf v})^t\in \mathbb{R} \oplus \mathbb{R}^2, \; \alpha\ge 0, \;\alpha^2 - \|\textbf{v}\|^2\ge 0 \}$, which is the closure of the future lightcone in $\cal M$. The two Jordan algebras $\mathcal{H}(2,\mathbb{R})$ and $\mathbb{R}\oplus\mathbb{R}^2$ are 
isomorphic, as Jordan algebras, via the following map: %of Jordan algebras: 
\begin{equation}\label{eq:chi}
\begin{array}{cccl}
	\chi: & \mathcal H(2,\mathbb R)  & \stackrel{\sim}{\longrightarrow} & \mathbb R\oplus \mathbb R^2  \\
	&  \begin{pmatrix}
		\alpha + v_1 & v_2 \\
		v_2 & \alpha - v_1
	\end{pmatrix}  & \longmapsto & \begin{pmatrix}\alpha\\v_1\\v_2\end{pmatrix}.%(\alpha ,  (v_1,v_2)^t). 
\end{array}
\end{equation}
Clearly  
$\chi(\overline{\mathcal H^+}(2,\mathbb R))=\overline{\mathcal L^+}$. Using the isomorphism $\chi$, we can interpret eq. \eqref{eq:fund} of color measurements as the action of a geometric transformation on $\cal M$. Given a chromatic state $\textbf{s}$ and an effect $\textbf{e}$ parameterized by the vectors $\textbf{v}_{\textbf{s}}=(s_1,s_2)$ and $\textbf{v}_{\textbf{e}}=(e_1/e_0,e_2/e_0)$, respectively, it is shown in \cite{BerthierProvenzi:2022PRS} that, whenever $||\textbf{v}_{\textbf{e}}||<1$ we have
%Let $\textbf{e}=(e_0,e_1,e_2)$ be an effect, with $\textbf{v}_{\textbf{e}}=(e_1/e_0,e_2/e_0)$ s.t. $||\textbf{v}_{\textbf{e}}||<1$ then
\begin{equation}\label{eq:Boost}
\chi(\psi_{\textbf{e}}(\textbf{s}))=\frac{e_0}{\gamma_{\textbf{v}_{\textbf{e}}}}B(\textbf{v}_{\textbf{e}})\frac{1}{2}\begin{pmatrix}1\\\textbf{v}_{\textbf{s}}\end{pmatrix}%=||\textbf{e}||_{\mathcal{M}}B(\textbf{v}_{\textbf{e}})\frac{1}{2}\begin{pmatrix}1\\\textbf{v}_{\textbf{s}}\end{pmatrix},
\end{equation}
where
\begin{equation}
\gamma_{\textbf{v}_{\textbf{e}}}= \frac{1}{\sqrt{1-||\textbf{v}_{\textbf{e}}||^2}}
\end{equation}
and the matrix representation of $B(\textbf{v}_{\textbf{e}})$ is given by%in the canonical basis is 
\begin{equation}\label{eq:boost}
[B(\textbf{v}_{\textbf{e}})]= \begin{pmatrix}
	\gamma_{\textbf{v}_{\textbf{e}}} & \gamma_{\textbf{v}_{\textbf{e}}}\textbf{v}_{\textbf{e}}^t \\ \gamma_{\textbf{v}_{\textbf{e}}}\textbf{v}_{\textbf{e}} & \sigma_0 + \frac{\gamma_{\textbf{v}_{\textbf{e}}}^2}{1+ \gamma_{\textbf{v}_{\textbf{e}}}}\textbf{v}_{\textbf{e}}\textbf{v}_{\textbf{e}}^t
\end{pmatrix},
\end{equation}
%The coefficient $e_0/\gamma_{\textbf{v}_{\textbf{e}}}=\sqrt{e_0^2-e_1^2-e_2^2}$ is the Minkowski norm $||\textbf{e}||_{\mathcal{M}}$ of the effect $\textbf{e}$. 
which can be recognized to be the Lorentz boost of $\cal M$ parameterized by the vector $\textbf{v}_{\textbf{e}}$.  

According to Minkowski geometry, this boost leaves the future lightcone invariant. The right-hand side of eq. (5) is the vector of $\overline{\mathcal L^+}$ that represents the outcome of the measurement given by eq. (3). Besides emphasizing the relativistic aspect of color perception, see also \cite{BerthierProvenzi:2022PRS,Berthier:21JMP}, this geometric description is very useful to analyze transformations between perceived color and to design white balance algorithms, as it will be discussed in the sequel.

The isomorphism $\chi$ provides a bridge between the two different perspectives on color measurements given by quantum information and Minkowski geometry. We underline here that eq. (3) expresses precisely the way in which an observer, represented by an effect, perceives a color of a visual scene prepared for a measurement in a given state. This description is radically different from the usual CIE color description because it relies on the very act of perceiving, and the geometric transformation in eq. \eqref{eq:Boost} corresponds actually to perceived color measurements.

\medskip

\section{Color measurements and split-quaternions}
\label{Split}
We describe now the last Jordan algebra $\mathbb S_0$ which is a sub-algebra of the split-quaternion algebra, and how to encode with simple algebraic equations the previous color measurement formulas by means of the so-called `sandwich formulas'.

\subsection{The algebra of split-quaternions}
The non-commutative, associative algebra of split-quaternions $\mathbb{S}$ is generally introduced in a similar way as the usual algebra of Hamilton's quaternions $\mathbb{H}$, see e.g. \cite{Gog:14,Gog:2022,Ozdemir:06} for more details. 
There are four basis elements, denoted by $1,i,j,k$,  however, differently from classic quaternions, $i$ and $j$ are such that $i^2=j^2=1$, furthermore $ij=-ji$. %=k$. 
%The four basis elements are $1,i,j,k$
%there is a basis composed by $1,i,j,k$, where the units $i$ and $j$ are such that $i^2=j^2=1$ and $ij=-ji$.
The element $k$ is defined as $k=ij$, this implies that $k^2=-1$. Moreover, the following multiplication rules hold: $kj = -jk = i$ and $ ik =-ki= j$.
% As a consequence it holds that $k^2=-1$.
%Using the properties introduced up to now it is easy to check the following multiplication rules:
%\begin{equation}
%ij = k, \quad jk = -i, \quad ki = -j, \quad ji = -k, \quad kj = i, \quad ik = j. 
%\end{equation}
%and one can obtain the following multiplication rules:
%\begin{equation}\label{eq:mult}
%\begin{split}
%ij &= k, \quad jk = -i, \quad ki = -j, \\ji &= -k, \quad kj = i, \quad ik = j.
%\end{split}
%\end{equation}
%Moreover the elements of the basis anti-commute so:
%\begin{equation}
%ji = -k, \quad kj = i, \quad ik = j. 
%\end{equation}
Every split-quaternion $q$ can be written as %$q = q_0+ q_1 i +q_2 j + q_3 k$, 
\begin{equation}\label{eq:form}
q = q_0+ q_1 i +q_2 j + q_3 k,  
\end{equation}
with $q_i\in \mathbb{R}$, $i\in \{0,\dots,3\}$. The real constant $q_0$ is called the scalar part of $q$, while ${\bf v}_q:=q_1 i +q_2 j + q_3 k$ is its vector part. The multiplication of two generic split-quaternions $q,r\in \mathbb{S}$ can be easily computed by writing them in the form of eq. \eqref{eq:form} and then using the multiplication rules described above, obtaining the following explicit expression:
\begin{equation}\label{eq:explicit_multiplication}
\begin{split}
	qr = &\quad q_0r_0 + q_1r_1 + q_2r_2 - q_3r_3  +  (q_0r_1 + q_1r_0 -q_2r_3 + q_3r_2)i\\ + & (q_0r_2 + q_2r_0 + q_1r_3 - q_3r_1)j  +  (q_0r_3+q_3r_0+q_1r_2-q_2r_1)k.
\end{split}
\end{equation}
%\begin{equation}\label{eq:explicit_multiplication}
%\begin{split}
%    qr = &\quad q_0r_0 + q_1r_1 + q_2r_2 - q_3r_3  \\  + & (q_0r_1 + q_1r_0 -q_2r_3 + q_3r_2)i\\ + & (q_0r_2 + q_2r_0 + q_1r_3 - q_3r_1)j \\  + & (q_0r_3+q_3r_0+q_1r_2-q_2r_1)k.
%\end{split}
%\end{equation}
In the same way as in the usual quaternion algebra $\mathbb{H}$, the conjugate $q^*$ of a split-quaternion $q$ % $q=q_0+q_1i+q_2j + q_3k$ 
is obtained by changing the sign of its vector part. %, thus:
%\begin{equation}
%    q^{*} = q_0 - q_1i - q_2j - q_3k.
%\end{equation}
The squared norm of a split-quaternion is given by $N^2(q)= qq^{*} = q_0^2 -q_1^2 -q_2^2 + q_3^2$.
%\begin{equation}
%N^2(q)= qq^{*} = q_0^2 -q_1^2 -q_2^2 + q_3^2.
%\end{equation}
Differently from $\mathbb{H},$ $N^2$ is not positive-definite and the split-quaternions are classified according to the sign of $N^2(q)$: if $N^2(q)<0$, $q$ is \textit{space-like}; if $N^2(q)=0$, $q$ is \textit{light-like} and if $N^2(q)>0$, $q$ is \textit{time-like}. 

We denote with $\mathbb S_s$, $\mathbb S_l$ and $\mathbb S_t$ the subsets of $\mathbb S$ containing space-like, light-like and time-like split quaternions, respectively. If $q\in \mathbb S_t$, then ${\bf v}_q$ can belong to both $\mathbb S_s$ and $\mathbb S_t$. However, if  $q\in \mathbb S_s$, then ${\bf v}_q$ can only belong to  $\mathbb S_s$. The first statement follows from the fact that, if $q\in \mathbb S_t$, then $q_0^2+q_3^2>q_1^2+q_2^2$, so both the cases $q_1^2+q_2^2<q_3^2$ or $q_3^2<q_1^2+q_2^2$ are possible, implying $N^2({\bf v}_q)=-q_1^2 - q_2^2 + q_3^2>0$ or $N^2({\bf v}_q)=-q_1^2 - q_2^2 + q_3^2<0$, i.e. ${\bf v}_q\in \mathbb S_t$ or ${\bf v}_q\in \mathbb S_s$, respectively.

To verify the last statement, consider $q\in \mathbb S_s$, then $q_0^2+q_3^2<q_1^2+q_2^2$, which implies $q_3^2<q_1^2+q_2^2$, thus $N^2({\bf v}_q)=-q_1^2 - q_2^2 + q_3^2<0$ and so ${\bf v}_q\in \mathbb S_s$.

Contrary to %the usual quaternion algebra
$\mathbb{H}$, $\mathbb{S}$ is not a division ring.
In fact, it is easy to show that %not all the elements of $\mathbb{S}$ admit a multiplicative inverse,
light-like split-quaternions do not admit a multiplicative inverse. The non-commutative, associative algebra $\mathbb{S}$ is isomorphic to the algebra of $2\times 2$ matrices with real entries via the following map
\begin{equation}\label{eq:zeta}
\begin{array}{cccl}
	\zeta: & \mathbb{S} & \stackrel{\sim}{\longrightarrow} & \mathcal M(2,\mathbb R)   \\
	&  q_0+ q_1 i +q_2 j + q_3 k   & \longmapsto & \begin{pmatrix}
		q_0 + q_1 & q_2 + q_3 \\
		q_2-q_3 & q_0 - q_1
	\end{pmatrix}.
\end{array}
\end{equation}
The split-quaternion multiplication corresponds to the matrix multiplication via $\zeta$, hence $\zeta(qr)=\zeta(q)\zeta(r)$, for all $q,r\in \mathbb{S}$.

\subsection{The sub-algebra $\mathbb S_0$ of the split-quaternion algebra}
Let us denote $\mathbb{S}_0$ the set of split-quaternions such that $q_3 = 0$, so every $q\in \mathbb S_0$ has the following form $q=q_0+q_1i+q_2j$. $\mathbb{S}_0$ becomes a commutative but not associative Jordan algebra when it is equipped with the Jordan product $q \circ r= (qr+rq)/2$. The restriction of $\zeta$ to $\mathbb{S}_0$, still denoted with $\zeta$ for simplicity, induces the following isomorphism of Jordan algebras:  
%\begin{equation}\label{eq:Jpr}
%    q \circ r= (qr+rq)/2%\frac{1}{2}(qr+rq)
%\end{equation}
\begin{equation}\label{eq:zeta}
\begin{array}{cccl}
	\zeta: & \mathbb{S}_0 & \stackrel{\sim}{\longrightarrow} & \mathcal H(2,\mathbb R)   \\
	&  q_0+ q_1 i +q_2 j  & \longmapsto & \begin{pmatrix}
		q_0 + q_1 & q_2 \\
		q_2 & q_0 - q_1
	\end{pmatrix}. 
\end{array}
\end{equation}
As a consequence, if $\overline{\mathbb{S}_0^+}$ indicates the domain of positivity of $\mathbb{S}_0$, then $\zeta(\overline{\mathbb{S}_0^+})=\overline{\mathcal H^+}(2,\mathbb R)$. 
If $\sigma_1$ and $\sigma_2$ denote the Pauli matrices with real entries, i.e.
\begin{equation}
\sigma_1 = \begin{pmatrix}
	1 & 0 \\ 0 & -1
\end{pmatrix}, \quad    \sigma_2 =  \begin{pmatrix}
	0 & 1 \\ 1 & 0
\end{pmatrix}
\end{equation}
then $\zeta^{-1}(\sigma_1)=i$, $\zeta^{-1}(\sigma_2)=j$.
%\begin{equation}
%\zeta^{-1}(\sigma_1)=i, \quad \zeta^{-1}(\sigma_2)=j.
%\end{equation}
The real Pauli matrices play a fundamental role in the quantum interpretation of color perception since they encode Hering's opponent mechanism, see e.g. \cite{Berthier:2020,BerthierProvenzi:2022PRS}. 
%Let us denote with $\sigma_0 =Id_2$, then it holds also that $\zeta^{-1}(\sigma_0)=1$.

The spin factor $\mathbb{R}\oplus \mathbb{R}^2$ is isomorphic, as a Jordan algebra, to the commutative and non-associative sub-algebra of the Clifford algebra $CL(2,Q)$, where $Q$ is the positive definite quadratic form on $\mathbb{R}^2$, linearly generated by the unit $1$ and a spin
system of $CL(2,Q)$. Let us recall that the Clifford algebra $CL(2,Q)$ is the quotient of the tensor algebra
\begin{equation}
T(\mathbb R^2)=\mathbb R\oplus\mathbb R^2\oplus (\mathbb R^2\otimes\mathbb R^2)\oplus\cdots=\bigoplus_{i\geq 0}(\mathbb R^2)^{\otimes i}
\end{equation}
by the two-sided ideal $\mathcal I(\mathbb R^2,Q)$ generated by the elements of the form $u\otimes u - Q(u)$, for $u$ in $\mathbb R^2$. For further details the interested reader can consult e.g. \cite{Postnikov:86}.

Since $CL(2,Q)$ is isomorphic to $\mathbb S$, this means that the map
\begin{equation}\label{eq:omeg}
\begin{array}{cccl}
	\omega: & \mathbb{S}_0 & \stackrel{\sim}{\longrightarrow} & \mathbb{R}\oplus \mathbb{R}^2   \\
	&  q_0+ q_1 i +q_2 j  & \longmapsto & (q_0,  q_1, q_2)^t,%\begin{pmatrix}
	%		q_0 & q_1 &q_2
	%	\end{pmatrix}^t,
%\begin{pmatrix}
%			q_0 \\ q_1 \\q_2
%		\end{pmatrix},
\end{array}
\end{equation}
is an isomorphism of Jordan algebras.

This completes the description of the three perspectives on color measurements listed in the introduction. The following commutative diagram of isomorphisms gives a concise mathematical representation of the relations between these perspectives:

\begin{equation}\label{eq:isoJA}
\begin{tikzcd}
{\mathcal{H}(2,\mathbb{R})}  \arrow[rr, "\chi"] &   & \mathbb{R}\oplus\mathbb{R}^2  \\
& \arrow[lu, "\zeta"']\mathbb{S}_0\arrow[ru, "\omega"'] & 
\end{tikzcd}
\end{equation}
According to what has been discussed before, this means that transformations corresponding to perceived color measurements can be computed in the split-quaternion framework.

\subsection{The sandwich formula for color measurements}

Our aim here is to express the original measurement equation (3), which was written in the setting of the Jordan algebra $\mathcal H(2,\mathbb R)$ and its geometric counterpart expressed by eq. (5), which exploited the isomorphism with the spin-factor $\mathbb R \oplus \mathbb R^2$, in terms of split-quaternions by making use of the Jordan algebra $\mathbb{S}_0$. 

To do so, we use the isomorphism $\zeta$ defined in eq. \eqref{eq:zeta} and the definitions \eqref{eq:rho} and \eqref{eq:effectmat} of density and effect matrix, respectively. It is clear that, if we define the two split quaternions $p_{\textbf{e}},q_{\textbf{s}}\in \overline{\mathbb{S}_0^+}$ as follows
\begin{equation}
p_{\textbf{e}} = e_0 + e_1i +e_2j, \quad q_{\textbf{s}} = %\frac{1}{2}
(1 + s_1i +s_2j)/2,
\end{equation}
then 
%\begin{equation}
%\begin{split}
%    \chi(\psi_{\textbf{e}}(\textbf{s}))&=\chi(\eta_{\textbf{e}}^{1/2}\rho_{\textbf{s}}\eta_{\textbf{e}}^{1/2})\\&=\chi(\zeta(p_{\textbf{e}}^{1/2})\zeta(q_{\textbf{s}})\zeta(p_{\textbf{e}}^{1/2}))\\
%    &= \chi(\zeta(p_{\textbf{e}}^{1/2}q_{\textbf{s}}p_{\textbf{e}}^{1/2})),
%\end{split}
%\end{equation}
\begin{equation}
\chi(\psi_{\textbf{e}}(\textbf{s}))=\chi(\eta_{\textbf{e}}^{1/2}\rho_{\textbf{s}}\eta_{\textbf{e}}^{1/2})=\chi(\zeta(p_{\textbf{e}}^{1/2})\zeta(q_{\textbf{s}})\zeta(p_{\textbf{e}}^{1/2}))= \chi(\zeta(p_{\textbf{e}}^{1/2}q_{\textbf{s}}p_{\textbf{e}}^{1/2})),
\end{equation}
%where $p_{\textbf{e}}\in$ and $q_{\textbf{s}}$ are the following split-%quaternions of $\overline{\mathbb{S}_0^+}$:
%\begin{equation}
%p_{\textbf{e}} = e_0 + e_1i +e_2j, \quad q_{\textbf{s}} = \frac{1}{2}(1 + s_1i +s_2j).
%\end{equation}
which, since $\chi\circ \zeta=\omega$, simplifies to 
\begin{equation}\label{eq:sandwich}
\chi(\psi_{\textbf{e}}(\textbf{s}))= \omega(p_{\textbf{e}}^{1/2}q_{\textbf{s}}p_{\textbf{e}}^{1/2}).
\end{equation}
Thus, in the Jordan algebra $\mathbb{S}_0$ the color measurement is expressed as a sandwich of split-quaternions with the same split-quaternion $p^{1/2}_{\textbf{e}}$ on both sides. We must stress that this is different than the conjugation action $q \mapsto \alpha q \alpha^{*}$, in which, figuratively speaking, one of the bread slices appears as conjugated. The consequences of this particular sandwich formula will be explained in detail in the following section.
%involving the split-quaternion conjugate. 

Given $\textbf{e}$, it is of course  
useful to know how to obtain an explicit expression for $p_{\textbf{e}}^{1/2}$.
Let us recall, see \cite{Gog:14}, %,Gog:2022},
that every time-like split-quaternion $p = p_0+p_1i+p_2j\in \mathbb{S}_0$ with a space-like vector part can be written as:
\begin{equation}\label{eq:polar}
p = N(p)U(p)= N(p)(\cosh\vartheta + u_p\sinh\vartheta),
\end{equation}
where $\cosh\vartheta = \frac{p_0}{N(p)}$ and $\sinh\vartheta= \frac{\sqrt{p_1^2+p_2^2}}{N(p)}$,
%\begin{equation}
%    \cosh\vartheta = \frac{p_0}{N(p)}, \quad\quad \sinh\vartheta= \frac{\sqrt{p_1^2+p_2^2}}{N(p)},
%\end{equation}
$u_p$ being the following unit space-like split-quaternion:
\begin{equation}
u_p = \frac{p_1i +p_2j}{\sqrt{p_1^2+p_2^2}}.
\end{equation}
Using this polar form it is easy to check that such a split-quaternion $p$ admits a unique square root, see \cite{Ozdemir:09}, given by
\begin{equation}
p^{1/2}=\sqrt{N(p)}(\cosh(\vartheta/2)+ u_p\sinh(\vartheta/2)).
\end{equation}
In our case, since $p_{\textbf{e}}= \zeta^{-1}(\eta_{\textbf{e}})= e_0 + e_1i + e_2j$, $p_{\textbf{e}}^{1/2}$ can be easily calculated knowing the coordinates of $\textbf{e}$, in fact:
\begin{equation}\label{eq:par}
\sqrt{N(p_{\textbf{e}})} = \sqrt[4]{e_0^2 - e_1^2 - e_2^2}, \qquad
\vartheta_{\textbf{e}}=\text{arctanh}||\textbf{v}_{\textbf{e}}||, \qquad
u_{p_{\textbf{e}}} = \frac{e_1i + e_2j}{\sqrt{e_1^2+e_2^2}}.
\end{equation}

\subsection{A sandwich without conjugate}
In this paragraph we are going to provide an explanation of the lack of conjugation in the split-quaternion sandwich formula of eq. \eqref{eq:sandwich}. To do so we start by recalling the geometric interpretation of the classic quaternion and split-quaternion conjugation formula used to obtain respectively classic rotations and Lorentz boosts.

It is well known that in classic quaternions the action of conjugation by a unit quaternion $\alpha$, i.e. $q \mapsto \alpha q\alpha^*$ for any $q\in\mathbb{H}$, is an efficient way of encoding rotations of $\mathbb{R}^3$, \cite{Kaufmann:05}.
%To understand why the conjugation action corresponds to a rotation 
Let us start by recalling that any rotation $R$ in $\mathbb{R}^3$  is fully determined by a rotation angle $\vartheta$ and a rotation axis $v\in \mathbb{R}^3$. Moreover, when written with respect to the orthogonal splitting of $\mathbb{R}^3$ given by $v,v^{\perp}$, its matrix is
\begin{equation}\label{eq:rotations}
R = \begin{pmatrix}
1 & 0 \\  0 & R_{\vartheta}
\end{pmatrix}, \quad\quad\quad \text{with } \quad  R_{\vartheta}= \begin{pmatrix}
\cos\vartheta & -\sin\vartheta \\ \sin\vartheta & \cos\vartheta
\end{pmatrix}.
\end{equation}
%where $R_{\vartheta}$ denotes the anticlockwise rotation matrix of angle $\vartheta$ on $v^{\perp}$ given by
%\begin{equation}
%    R_{\vartheta}= \begin{pmatrix}
%        \cos\vartheta & -\sin\vartheta \\ \sin\vartheta & \cos\vartheta
%    \end{pmatrix}.\end{equation}
Let us now identify $\mathbb{H}$ with $\mathbb{R}^4$ as follows:
\begin{equation}\label{eq:iota}
\begin{array}{cccl}
\iota: & \mathbb{H} & \stackrel{}{\longrightarrow} & \mathbb{R}^4   \\
&  q_0+ q_1 i +q_2 j + q_3 k & \longmapsto & (q_0,  q_1, q_2,q_3)^t.	\end{array}
\end{equation}
Any unit quaternion $\alpha$ can be written in the following form:
\begin{equation}\alpha = \cos(\vartheta/2) + u\sin(\vartheta/2),
\end{equation}
with $u$ being a unit quaternion in $\text{span}(i,j,k)$, \cite{Gog:14}. We want to show that the conjugation action by $\alpha$ of a quaternion $q$, i.e. $q \mapsto \alpha q\alpha^*$, corresponds, to a rotation of axis $v=\iota(u)$ and angle $\vartheta$ in the three-dimensional vector subspace $V=\iota(\text{span}(i,j,k))$ %=\iota(\text{span}(u,u^{\perp}))$ 
of $\mathbb{R}^4$. %In literature, this result is known under the name of Rodrigues quaternionic formula.

%$V\cong \mathbb{R}^3$, defined as $V=\iota(\text{span}(i,j,k))=\iota(\text{span}(u,u^{\perp}))\subset\mathbb{R}^4$.

Writing the elements of $\mathbb{H}$ w.r.t. the basis $1, u, u^{\perp}$, after straightforward computations, one can see that the left multiplication by $\alpha$ of a quaternion $q$ in $\mathbb{R}^4$ corresponds to:
\begin{equation}\label{eq:left}
\iota(\alpha q) = \begin{pmatrix}
R_{\vartheta/2} & O_2 \\ O_2 & R_{\vartheta/2}
\end{pmatrix}\iota(q),
\end{equation}
where $O_2$ refers to the $2 \times 2$ null matrix. Analogously, the right multiplication of $q$ by its conjugate $\alpha^*=\alpha^{-1}$, gives:
\begin{equation}\label{eq:right}
\iota(q\alpha^*) = \begin{pmatrix}
R_{-\vartheta/2} & O_2 \\ O_2 & R_{\vartheta/2}
\end{pmatrix}\iota(q).
\end{equation}
Combining eq. \eqref{eq:left} and \eqref{eq:right}, one obtains the following matrix:
\begin{equation}\label{eq:rotconj}
\iota(\alpha q\alpha^*) = \begin{pmatrix}
Id_2 & O_2 \\ O_2 & R_{\vartheta}
\end{pmatrix}\iota(q).
\end{equation}
Comparing the latter expression with eq. \eqref{eq:rotations} it is clear that $\left. \iota(\alpha q\alpha^*)\right|_V=R\iota(q)$, hence we can conclude that the conjugation action corresponds to a rotation in $V$.

\bigskip

Let us consider now the case of split-quaternions and Lorentz boosts. Let us interpret $\mathbb{R}^3$ as the three-dimensional Minkowski space-time and express it using the basis $x,y,t$, in which the first two elements are the spatial dimensions and the latter represents the temporal one. A Lorentz boost $H$ in $\mathbb{R}^3$ is determined by an axis of fixed points $v\in\text{span}(x,y)$ and an angle of hyperbolic rotation $\vartheta$, called \textit{boost rapidity}. Let us indicate with $z$ the orthogonal vector to $v$ in $\text{span}(x,y)$, sometimes in the literature $z$ is called \textit{boost direction}. Expressing the boost matrix w.r.t. the basis $v,z,t$, one obtains the following analogue of eq. \eqref{eq:rotations}, with a hyperbolic rotation instead of a rotation:
\begin{equation}\label{eq:hyprotations}
H = \begin{pmatrix}
1 & 0 \\  0 & H_{\vartheta}
\end{pmatrix}, \quad\quad\quad \text{with } \quad  H_{\vartheta}=\begin{pmatrix}
\cosh\vartheta & \sinh\vartheta\\ \sinh\vartheta & \cosh\vartheta
\end{pmatrix}.
\end{equation}
Notice that the hyperbolic rotation $H_\vartheta$ occurs on the vector subspace $\text{span}(z,t)$ involving the boost direction and the time axis. Moreover $H_{\vartheta}^{-1}=H_{-\vartheta}$ and $H_{\vartheta_1}H_{\vartheta_2}=H_{\vartheta_1+\vartheta_2}$.

Now let us identify the split-quaternion algebra $\mathbb{S}$ with $\mathbb{R}^4$ via the function $\iota$ of eq. \eqref{eq:iota}, $\iota: \mathbb{S} \longrightarrow \mathbb{R}^4$. As before, we define the vector subspace $V=\iota(\text{span}(i,j,k))$. Our aim is to show that the conjugation action by %a unit split-quaternion
%time-like split-quaternion (with space-like vector part) 
$\alpha$ corresponds to a Lorentz boost in $V$. Notice that, in this identification, $\iota(k)=t$ plays the role of the time axis, while $\iota(\text{span}(i,j))=\text{span}(x,y)$ represents the 2-dimensional space.

Let us consider a unit time-like split-quaternion $\alpha$ with space-like vector part. By eq. \eqref{eq:polar}, $\alpha$ can be written as follows:
\begin{equation}
\alpha = \cosh(\vartheta/2) + u\sinh(\vartheta/2),
\end{equation}
with $u = u_1 i +u_2 j$, being a unit split-quaternion. Let $w = -u_2 i + u_1j$ be the vector orthogonal to $u$ in $\text{span}(i,j)$.
As before, expressing the split-quaternions of $\mathbb{S}$ w.r.t. the basis $1, u, w, k$ and recalling the split-quaternion multiplication of eq. \eqref{eq:explicit_multiplication}, we obtain that the left and right multiplication, respectively, of any $q\in\mathbb{S}$ by $\alpha$ in $\mathbb{R}^4$ are given by:

\begin{equation}\label{eq:hleftright}
\iota(\alpha q) =  \begin{pmatrix}H_{\vartheta/2} & O_2 \\ O_2  & H_{-\vartheta/2}
\end{pmatrix} \iota(q), \qquad \iota(q\alpha) =  \begin{pmatrix}H_{\vartheta/2} & O_2 \\ O_2  & H_{\vartheta/2}
\end{pmatrix} \iota(q).
\end{equation}

Since $\alpha$ is not a light-like split-quaternion, it admits a multiplicative inverse, and since we have supposed it to be a unit split quaternion, $\alpha^{-1}=\alpha^*$, see e.g. \cite{Gog:14}. Thus the left multiplication by $\alpha^*$ is given by:
\begin{equation}\label{eq:hconjleft}
\iota(\alpha^* q) =  \begin{pmatrix}H_{-\vartheta/2} & O_2 \\ O_2  & H_{\vartheta/2}
\end{pmatrix} \iota(q).
\end{equation}
Similarly to what was done in eq. \eqref{eq:rotconj}, we can combine eqs. \eqref{eq:hleftright} and \eqref{eq:hconjleft} obtaining:
\begin{equation}\label{eq:sand}
\iota(\alpha^*q\alpha) =  \begin{pmatrix}Id_2 & O_2 \\ O_2  & H_\vartheta
\end{pmatrix} \iota(q).
\end{equation}
Comparing the last equation with eq. \eqref{eq:hyprotations} one can see that $\left.\iota(\alpha^* q\alpha)\right|_V=H\iota(q)$, thus the conjugation action $q \mapsto \alpha^*q\alpha$ corresponds to a Lorentz boost in $V$ of axis $\iota(u)$, direction $\iota(w)$ and rapidity $\vartheta$. We must stress that the most commonly used formula (e.g. for the change of reference frames in relativity), % and in the application described in the next section)
is actually the inverse of the previous one, i.e. $q \mapsto \alpha q\alpha^*$ obtained changing the sign of the rapidity.

\bigskip

We have just seen how the conjugation action $\alpha \mapsto \alpha^*q\alpha$ corresponds to a boost in the vector subspace $V=\iota(\text{span}(i,j,k))=\iota(\text{span}(u,w,k))$, in which the time coordinate is associated to the split-quaternion $k$ via $\iota$. However, as discussed in the previous paragraphs, we are interested in the sub-algebra $\mathbb{S}_0$ of split-quaternions having $q_3=0$ in which, recalling the isomorphism $\omega$ of eq. \eqref{eq:omeg}, time is associated to $1$ instead of $k$.

It is convenient to introduce the vector subspace $W = \iota(\mathbb{S}_0)=\iota(\text{span}(1,i,j))$.
For coherence with the relativistic interpretation of the model, we would like a split-quaternion sandwich formula corresponding to a Lorentz boost in $W$, instead of $V$.

Let us finally analyze the sandwich formula $q \mapsto \alpha q \alpha$, which is the one that will be used to implement the colorimetric transformations that will be discussed in the following section. Using the expressions in eq. \eqref{eq:sandwich} one obtains that this sandwich without conjugate, in $\mathbb{R}^4$, corresponds to:
\begin{equation}\label{eq:sandnocon}
\iota(\alpha q\alpha) =  \begin{pmatrix} H_\vartheta& O_2 \\ O_2  & Id_2\end{pmatrix} \iota(q),
\end{equation}
where the matrix above is written w.r.t. the basis $\iota(1), \iota(u), \iota(w), \iota(k)$, hence it represents a boost in $W$ of axis $\iota(w)$, direction $\iota(u)$ and rapidity $\vartheta$.

Notice that this is different from the case of eq. \eqref{eq:sand}, in which $\iota(u)$ was the boost axis and $\iota(w)$ its direction. Nevertheless, in both cases $\alpha$ is parameterized by $u$, thus we must stress that when using eq. \eqref{eq:sandnocon} we pass as parameters, contained in $\alpha$, the information associated to the boost direction and rapidity. This will be useful in the following section, since we associate the information of the illuminant vector to the boost direction and rapidity, that will be given directly as input parameters contained in $\alpha$.

%As we will see in the following section this is also useful because we give as input directly parameters concerning the illuminant vector,, without the need of calculating its orthogonal, that would be necessary if one would wanted to use the formulation of eq.
%To have a normalized boost $\alpha = U(p_{\textbf{e}}^{1/2})$, either we multiply $\alpha$ by $\sqrt{N(p_{\textbf{e}})}$, either everything by $N(p_{\textbf{e}})$.
Note that $\left.\iota(\alpha q \alpha)\right|_W$ corresponds to the matrix in eq. \eqref{eq:boost} after a suitable change of basis.
Finally, let us recall that $p_{\textbf{e}}^{1/2}$ in eq. \eqref{eq:sandwich} is not a unit split-quaternion, hence, to use eq. \eqref{eq:sandnocon}, one must identify $\alpha = U(p_{\textbf{e}}^{1/2})$, and multiply eq. \eqref{eq:sandnocon} by $N(p_{\textbf{e}})$.

\medskip

\section{Application to white balance}\label{sec:experiments}
In this section we use the color measurements modeled in the context of split-quaternions to implement a novel chromatic adaptation transform (CAT) for the automatic white balance (AWB) of digital images. AWB is a classic color processing meant to let the camera mimic the adaptation of the human visual system to the chromaticity of the illumination condition, often called illuminant. It consists of two steps: an illuminant estimation part which identifies the illuminant(s) present in the visual scene, associating to them a 3-dimensional vector $L$, and a CAT, parametrized by $L$, returning an image representing how the scene would appear to an observer fully adapted to the illuminant. %as the scene was lit by a neutral illuminant.
Several CATs have been proposed in the literature, see e.g. \cite{CIEreport:04} for an overview, the most widely used for applications is the von Kries CAT, see \cite{vonKries:1902}.
We propose a perceptual CAT, called \textit{split-CAT} from now on, based on color measurements expressed using split-quaternions. A preliminary version of the algorithm, using simple Lorentz boosts, has been proposed in \cite{Guennec:21}, a more complete version, using normalized Lorentz boosts, can be found in chap. 7 of \cite{Prencipe:thesis}.

\begin{figure}[htbp]
\centering 
\includegraphics[scale = 0.345]{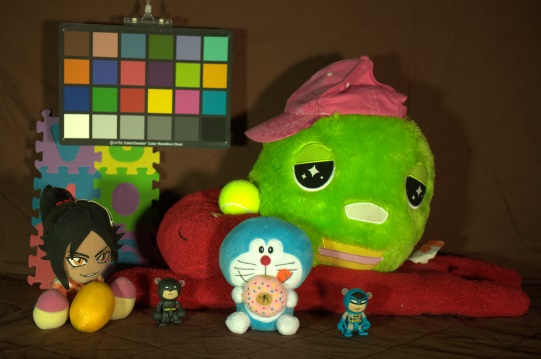}
\includegraphics[scale = 0.345]{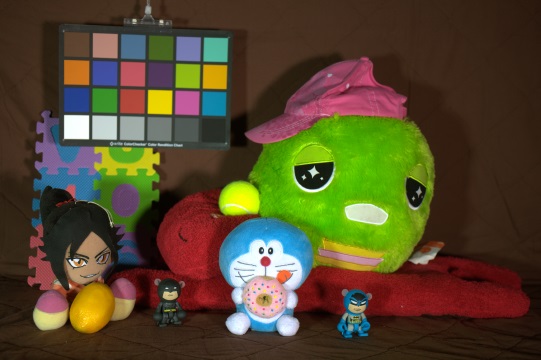}
\includegraphics[scale = 0.345]{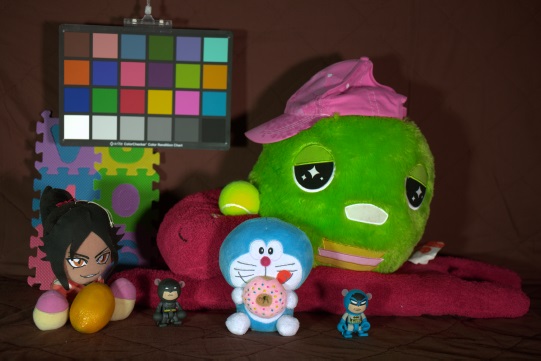}
\newline
\vspace{-0.3cm}

\hspace{-0.935cm}
\includegraphics[scale = 0.349]{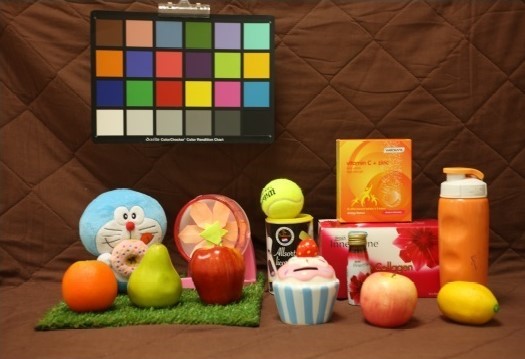}
\includegraphics[scale = 0.0279]{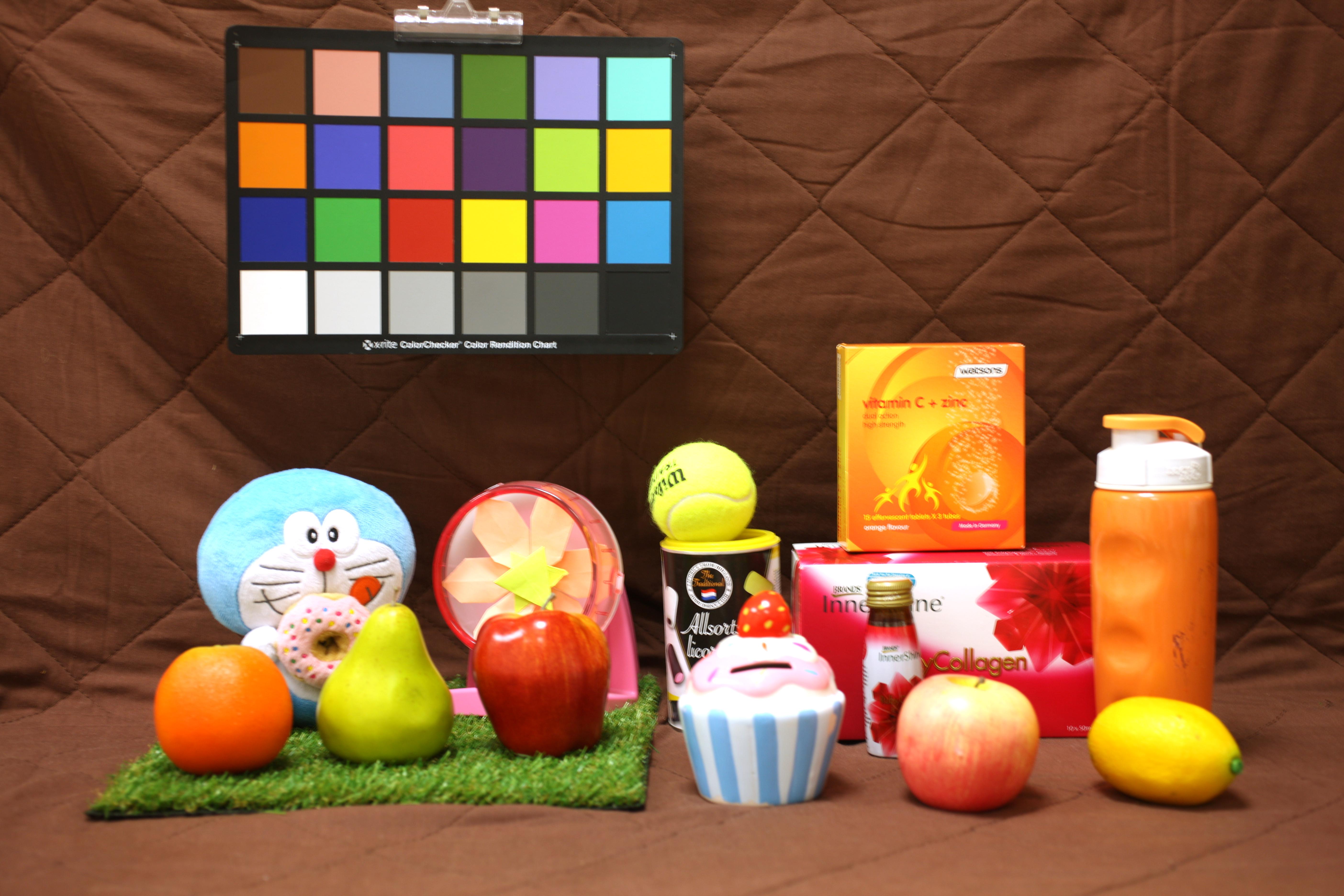}
\includegraphics[scale = 0.0279]{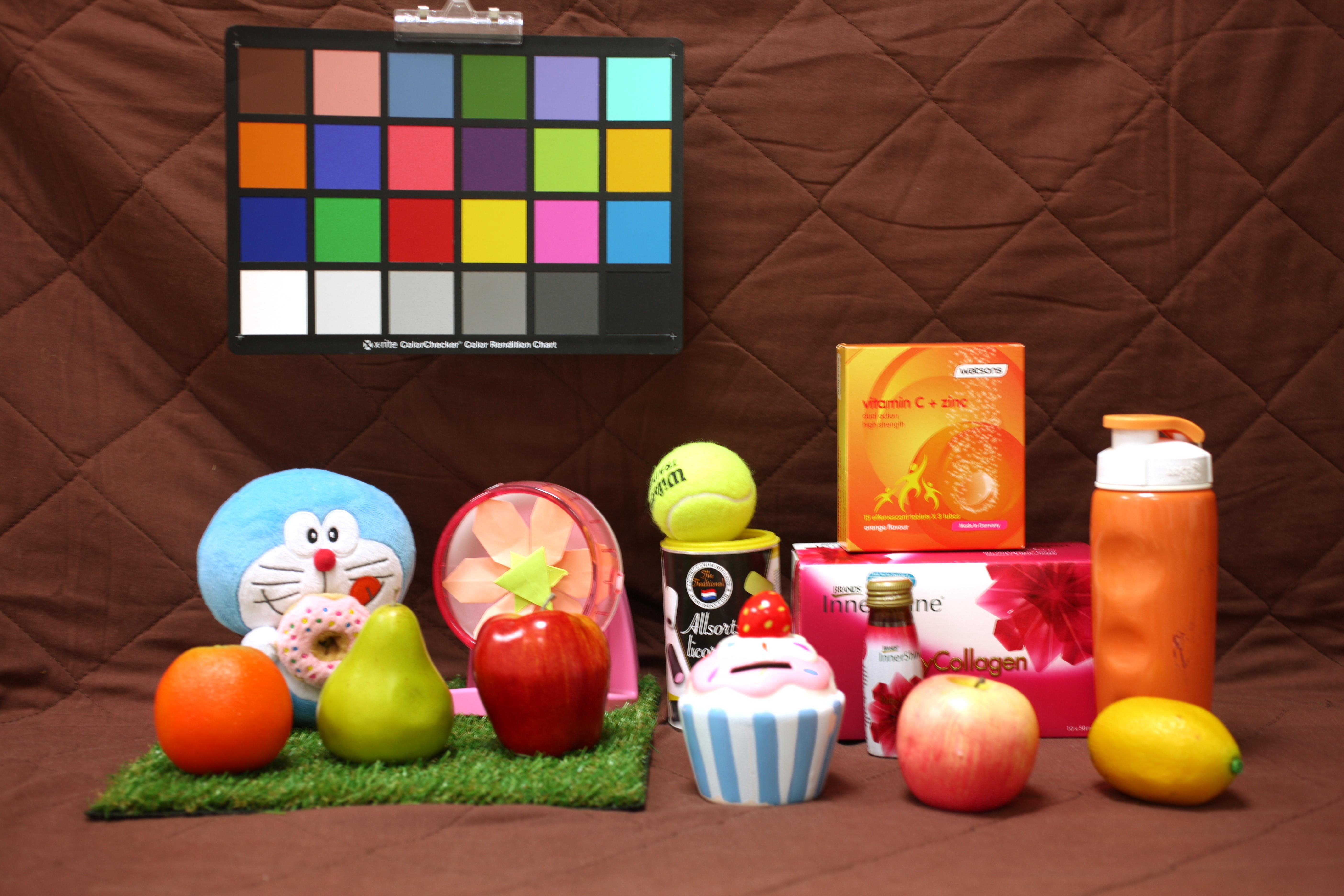}
\caption{\textit{Left}: input images. \textit{Center}: output images after white balance using the von Kries CAT. \textit{Right}: output images after white balance using the split-CAT. The white balanced images have been obtained using the same illuminant estimation. The illuminant vector is extracted automatically from the white patch of the color checker present in each image.}
\label{fig:imgs}
\end{figure}

Let us start by representing the input image in the split-quaternion domain $\overline{\mathbb{S}_0^+}$. We consider an input image $I$ with spatial domain denoted with $\mathcal{I}$, represented in the HCV color solid, thus $I(x)=(H(x),C(x),V(x))$, $\forall x \in \mathcal{I}$. To every pixel $x\in\mathcal{I}$, we associate the following split-quaternion $q(x)$ of $\overline{\mathbb{S}_0^+}$:
\begin{equation}
q(x) = V(x) + C(x)\cos(H(x)) i + C(x)\sin(H(x)) j.
\end{equation}
Given the output $L$ of an illuminant estimation algorithm, expressed in HCV color coordinates $(H_L,C_L,V_L)$, we associate to it the effect $\textbf{e}=(e_0,e_1,e_2)=(V_L,C_L\cos H_L, C_L\sin H_L)$. Then, the white balanced image $q'$ represented in $\overline{\mathbb{S}_0^+}$ is obtained using the split-quaternion multiplication as follows:
\begin{equation}
q'(x) = p^{-1/2}_{\textbf{e}} q(x) p^{-1/2}_{\textbf{e}}, \quad \forall x\in \mathcal{I},
\end{equation}
in which $p^{-1/2}_{\textbf{e}}$ is given by
\begin{equation}
p^{-1/2}_{\textbf{e}}=\frac{1}{\sqrt{N(p_{\textbf{e}})})}(\cosh(\vartheta_{\textbf{e}}/2)- u_{p_{\textbf{e}}}\sinh(\vartheta_{\textbf{e}}/2)),
\end{equation}
where $\sqrt{N(p_{\textbf{e}})}$, $\vartheta_{\textbf{e}}$ and $u_{p_{\textbf{e}}}$ are calculated from $\textbf{e}$ using eq. \eqref{eq:par}. Finally, we convert the image from $\overline{\mathbb{S}_0^+}$ back to the HCV color space. Some examples of outputs, obtained by processing images from the NUS Indoor Dataset \cite{Cheng:15}, are depicted in Figure \ref{fig:imgs}.

The results of the von Kries CAT and those of the split-CAT look very similar at first glance. This is already quite remarkable, because it gives a first concrete proof of the fact that a transformation predicted \textit{solely} by the rigorous mathematical interpretation of quantum measurement  within the quantum information model of color perception leads to a color transformation that produces results that are very similar to those of the most standard CAT. However, some hue shifts\footnote{The most visible case is the one of red objects shifting towards magenta.} can be seen in the output of the split-CAT, as shown by the hat in the third picture, first row, of Figure \ref{fig:imgs}. In the following subsection we will stress that this is not an intrinsic feature of the split-CAT, but it depends on the choice of the HCV color space and we will propose how to reduce the hue shifts by modifying this color space.

\subsection{Modification of HCV to encode Hering's opponency}
\begin{figure}[]
\centering 
\includegraphics[scale = 0.4]{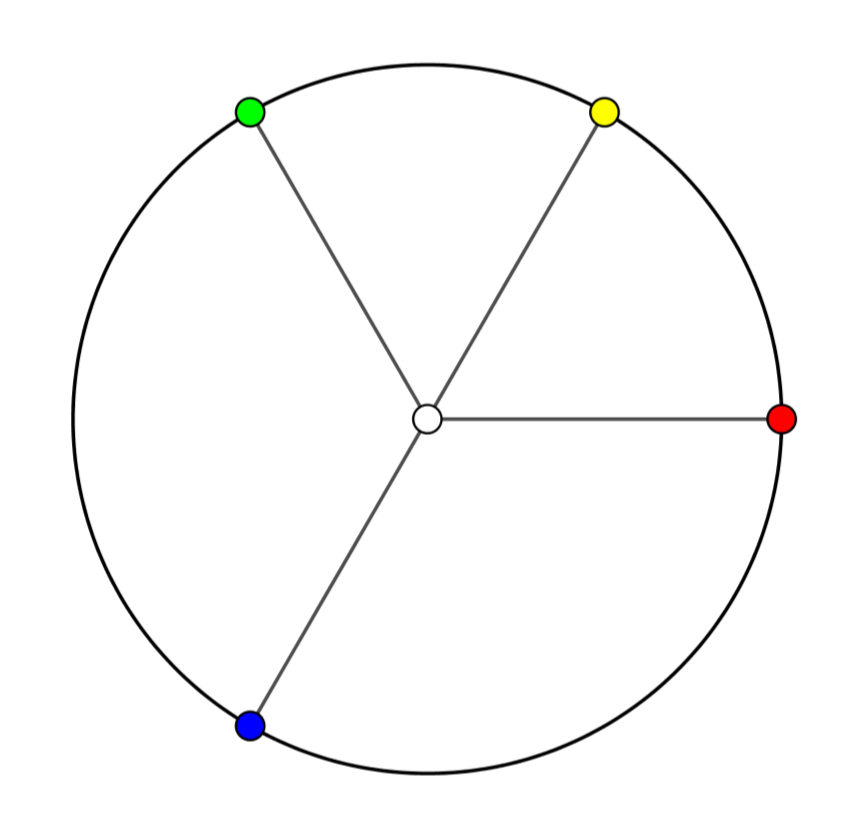}
\hspace{2.2mm}
\includegraphics[scale = 0.4]{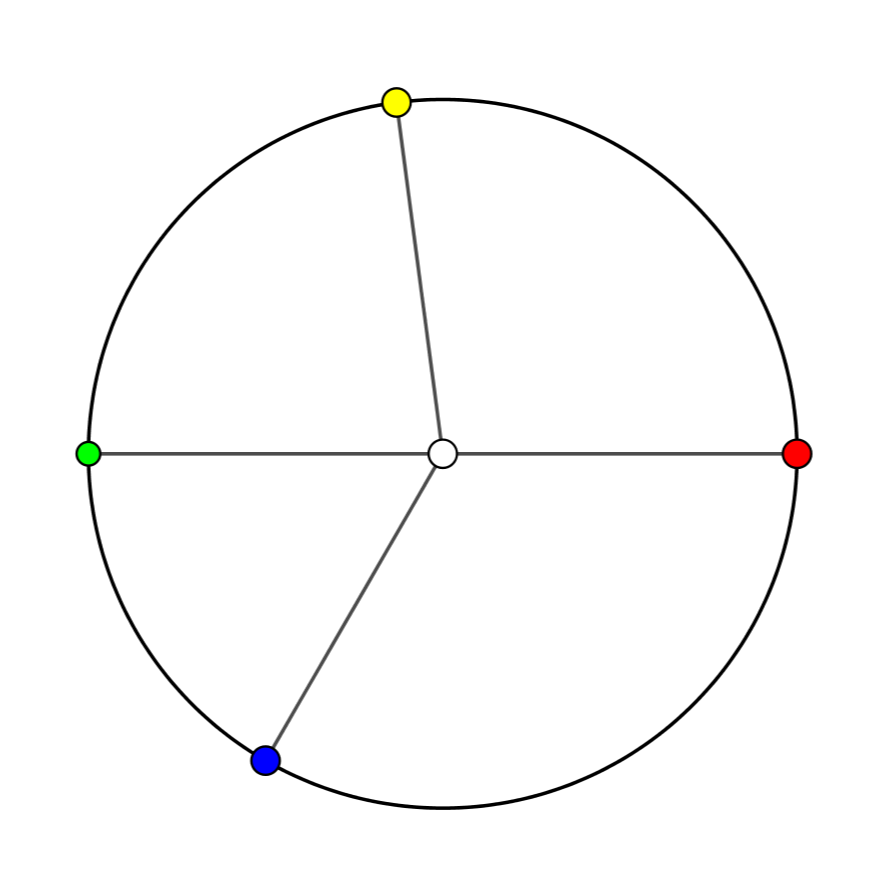}
\hspace{2mm}
\includegraphics[scale = 0.4]{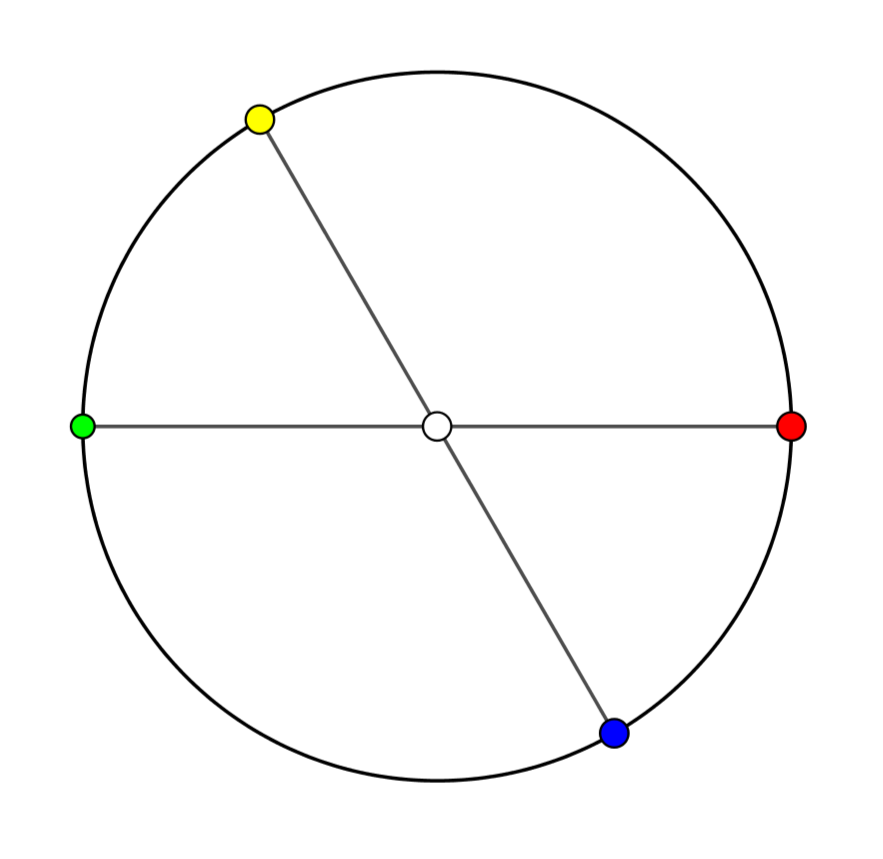}
\caption{Red, yellow, green and blue hue positions in the hue-chroma planes of the HCV, H$_1$CV and H$_2$CV color spaces.}
\label{fig:huepositions}
\end{figure}

The classic HCV color solid seemed to be a convenient approximation of the closed cone $\overline{\mathbb{S}^+_0}$ because of its conical shape. Nevertheless, HCV lacks of Hering's opponency, which instead is an intrinsic feature of the quantum information-based model of color perception. In fact, while blue and yellow are diametrically opposed in its hue configuration, it is not the case for red and green\footnote{Note that by yellow, blue, red and green here we mean the representations in the HCV color solid of the RGB vectors $(255,255,0)$, $(0,0,255)$, $(255,0,0)$, $(0,255,0)$. As remarked before there is no exact, nor clear, correspondence between these vectors and Hering's opponent hues, or unique hues.}, as it can be seen in the first picture from the left of Figure \ref{fig:huepositions}. Moreover, we were noticing that red objects were slightly turning pinkish after applying the split-CAT in HCV. For these reasons, we propose two modified versions of the HCV color domain, denoted with H${ }_1$CV and H${ }_2$CV, obtained from HCV by modifying its $H$ coordinate only. The objective is to modify the hue configuration on the circle in order to approximately recover the, non necessarily orthogonal, Hering's opponent axes. We must stress that a more general open issue in colorimetry is to understand which are the exact opponent unique hues and whether the opponent axes are orthogonal. Inter-observer variability and dependence on the viewing conditions clearly contribute to make the problem highly nontrivial.%, as depicted in fig. \ref{fig:huepositions}.

\begin{figure}[]
\centering 
\includegraphics[scale = 0.345]{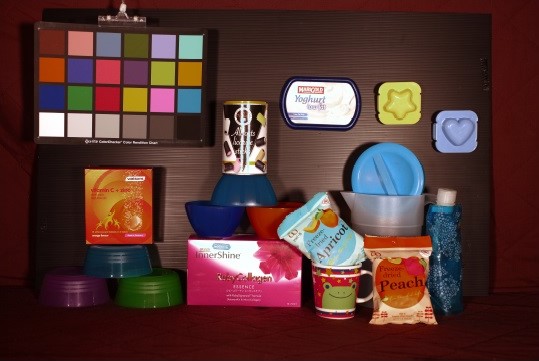}
\includegraphics[scale = 0.345]{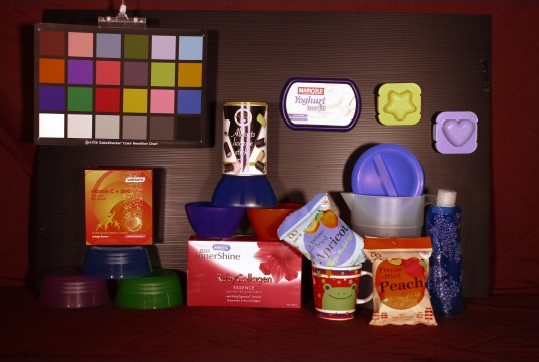}
\includegraphics[scale = 0.345]{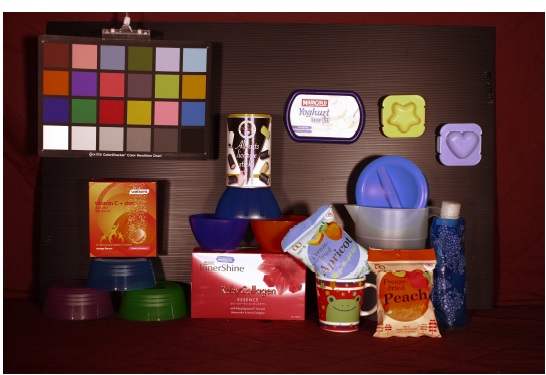}
\caption{\textit{Left}: output of the split-CAT in HCV. \textit{Center}: output of the split-CAT in H$_1$CV. \textit{Right}: output of the split-CAT in H$_2$CV.}%These images have been obtained from \cite{Cheng:15}.}
\label{fig:addingHering2}
\end{figure}

%and even whether it is possible to find them, since clearly inter-observer variability and dependence on the viewing conditions play an important role.
The split-CAT implemented in H${ }_1$CV or H${ }_2$CV gives overall better qualitative and quantitative, as we will see in the next paragraph, results. Figure  \ref{fig:addingHering2} shows an example\footnote{Notice that in particular does not produce a magenta shift of red objects, see e.g. the white-red box with a flower depicted on it.} obtained processing images from \cite{Cheng:15}.

Let us explain a bit more in detail how the color solids H${ }_1$CV and H${ }_2$CV have been constructed. We modified the $H$ coordinate trying several functions, obtained using simple interpolation techniques in 1-dimension and selected the two best performing ones on the images in the rendering of the red hue, let us call them $f_1, f_2$. Both $f_1,f_2:[0,2\pi]\rightarrow [0,2\pi]$ are $2\pi$-periodic and invertible. The coordinates H$_i$ %to change color space 
are obtained from $H$ by H$_i = f_i^{-1}(\text{H})$, $i=1,2$. In particular:
\begin{enumerate}
\item $f_1$ is obtained requiring the red to stay fixed, and the green to be diametrically opposed to the red, hence it is obtained by quadratic interpolation of the points $(0,0)$, $(2\pi/3,\pi)$, $(2\pi,2\pi)$. It can be explicitly written as a parabola $f_1(x)=\frac{1}{4}\left(7x-\frac{3}{2\pi}x^2\right)$. As depicted in Figure \ref{fig:huepositions} (\textit{Center}), red and green are now opponent, but the blue is diametrically opposed to an orangish yellow. Furthermore these opponent axes are not orthogonal, but separated by an angle of $30^\circ$.
%$$(0,0), \qquad \left(\frac{2\pi}{3},\pi\right), \qquad (2\pi,2\pi).$$
\item $f_2$ is obtained by fixing again the red and moving the green to be diametrically opposed to it, then moving the yellow and the blue in order to have an angle of $60^\circ$ between the two opponent axes, as in Figure \ref{fig:huepositions} (\textit{Right}). $f_2$ was obtained via quadratic piece-wise interpolation of the points $(0,0)$, $(\pi/3,2\pi/3)$, $(2\pi/3,\pi)$, $(4\pi/3,5\pi/3)$, $(2\pi,2\pi)$.
\end{enumerate}

\subsection{Quantitative evaluation of the color checker rendering}
We evaluated four CATs: the von Kries CAT, the split-CAT implemented in HCV, H$_1$CV and H$_2$CV. 

We started by generating linear PNG images applying linear demosaicing on the RAW RGB images provided with the NUS Indoor dataset (Canon 1Ds Mark III, 105 images), \cite{Cheng:15}. We automatically detected the color checker present in each image and extracted the nineteenth patch (the white one) as ground truth illuminant vector. Using the extracted ground truths we corrected the mentioned PNG images using the four different CATs. 

Clipping cases were managed by dividing the image by its maximum. Then we detected all the color checkers (a RGB vector corresponding to each patch) in the output images, still using the automatic color checker detection functions.

For each CAT we considered the set of detected color checkers and calculated the distance between each of them and the standard benchmark color checker, enlightened by the D65 illuminant. As distance we used seven state-of-the art color metrics listed in table \ref{tab:absolute distances2}. This distance was obtained by calculating, for each patch, its distance from the corresponding one in the benchmark color checker and then averaging over the 24 patches. For
each CAT we averaged the distances of the color checkers over the 105 images of the dataset,
obtaining the values reported by table \ref{tab:absolute distances2}. Both the algorithm for the automatic detection of the color checker and the different color metrics were used as implemented in the open-source Python package \textit{Colour}\footnote{See \href{https://colour.readthedocs.io/en/develop/}{https://colour.readthedocs.io/en/develop/}}.

\begin{table}[htbp]
\centering
\begin{tabular}{|l|c|c|c|c|}
\hline
Metrics & von Kries & split HCV& split H$_1$CV& split H$_2$CV\\
\hline
%		CIE 1994 \tiny{(textiles=T)} & 13.88 & \textbf{13.75} & 14.23& 14.04\\
CIE 1994 & 25.66 & 25.01 & 24.86 &\textbf{24.85}\\
%		DIN99 \tiny{(textiles=T)}  & 51.15 & 49.41 & \textbf{48.34} &48.66\\
DIN99   & 26.25 & 25.48 & \textbf{25.28} &25.39\\
CIEDE 2000 & 22.53 & \textbf{22.10} & 22.44&22.41\\
%		CAM02 SCD & 21.06 & \textbf{20.53} & 20.65&20.65\\
CAM02 UCS & 25.99 & \textbf{25.31} & \textbf{25.31}&25.34\\
CAM02 LCD  &  33.87& 32.99 & \textbf{32.84}&32.93\\
%		CAM16 SCD  & 21.06 & \textbf{20.53} & 20.65&20.66\\
CAM16 UCS  & 26.01 & \textbf{25.28} & 25.31 &25.35\\
CAM16 LCD  & 33.87 & 32.99 & \textbf{32.84}&32.94\\
\hline
\end{tabular}
\caption{Comparison between the von Kries CAT algorithm and split-CAT implemented in different HCV spaces.}
%	\caption{Average distances from the stan, von Kries CAT, normalized Lorentz boost in HCV,  normalized Lorentz boost in $H_1CV$, normalized Lorentz boost in $H_2CV$.}
\label{tab:absolute distances2}
\end{table}
Lower values in this table mean that the color checker rendering of a certain CAT is closer to the benchmark color checker. We can see that H$_1$CV is better performing than H$_2$CV. Furthermore, according to this evaluation, it is slightly better to use the implementation in H$_1$CV than in HCV. The value in bold highlights the smallest among the four values in the line in which it belongs, it can be seen that the three slip-CATs perform better than the von Kries CAT w.r.t. this criterion.

\section{Conclusions and future perspectives}
In this paper we have proposed and tested the first concrete application of a recently developed quantum-like framework for color perception. This application consists at defining a theoretically-based chromatic adaptation transform, the split-CAT, to perform the classic color processing task of automatic white balance. To do so, we started by enriching the algebraic structure of the quantum-like model by providing an alternative, isomorphic representation of the Jordan algebra $\mathcal{A}$, i.e. the sub-algebra $\mathbb{S}_0$ of the split-quaternion algebra $\mathbb{S}$, with $\overline{\mathbb{S}_0^+}$ as domain of positivity. 

In terms of split-quaternions, the fundamental measurement equation, eq. \eqref{eq:fund}, used in \cite{BerthierProvenzi:2022PRS,Berthier:22} to define a perceived color as the outcome of a measurement procedure, is expressed by a so-called sandwich formula, see eq. \eqref{eq:sandwich}. This sandwich formula is different from the one usually used  classically, since none of the terms appears conjugated. We better clarified this peculiarity by writing explicitly the action of the sandwich formula in $\mathbb{R}^4$.

Finally, we defined the split-CAT as the action of a sandwich, parametrized by the effect associated to the illuminant. To do so, we associated the pixels of a digital image represented in the HCV color solid to elements of $\overline{\mathbb{S}_0^+}$.

Qualitative considerations about the color rendering of the output images led to the definitions of two modified versions of the HCV color solid, H$_1$CV and H$_2$CV, obtained via simple interpolation techniques meant to modify the hue configuration, making it approximately closer to having Hering-like opponent axes. Then, we provided a quantitative evaluation of the performance of the classic von Kries CAT, the split-CAT in HCV, H$_1$CV and H$_2$CV in the overall rendering of the color checker patches. The results show that H$_1$CV provides the best score.

Further evaluations of the performance of the proposed algorithm are clearly possible, e.g. one could perform systematic comparisons with other state-of-the art CATs, using as a reference the corresponding colors datasets, as in \cite{CIEreport:04}.

An important open question concerns the domain for the implementation. A possible approach is to continuing fine-tuning the HCV color domain, or use other existing color spaces, like CIELab or IPT, keeping in mind that the algorithm is designed to preserve a conic shaped solid, hence color solids having more irregular shapes will probably produce artifacts that should be treated separately.

A different, interesting, perspective concerns the idea of studying alternative ways of integrating Hering's opponent mechanism in a color solid, which are more related to psychophysical data, see e.g. \cite{JamesonHurvichII:55,JamesonHurvichIII:56}, rather than performing a purely engineering-oriented a posteriori manipulation of existing color solids.

\section*{Acknowledgments}
This work has been partially carried out with the financial support of Huawei Technologies
France SASU and the European Research Council Advanced Grant (ERC AdG, ILLUSIVE: Foundations of Perception Engineering, 101020977).

\bibliographystyle{plain} 
\bibliography{bibliography}

\end{document}